\documentclass[12pt]{article}
\usepackage{amsmath}
\usepackage{latexsym}
\usepackage{bbm}
\usepackage{amssymb}

\def\a{\alpha}

\def\b{\beta}

\def\d{\delta}

\def\e{\varepsilon}

\def\z{\zeta}

\def\th{\theta}

\def\l{\lambda}

\def\o{\omega}

\def\s{\sigma}

\def\td{\theta^{\dagger}}

\def\ed{\e^{\dagger}}

\def\Yd{Y^{-1}}
\def\Zd{Z^{-1}}

\def\G{\Gamma}

\def\S{\Sigma}

\def\L{\Lambda}
\def\O{\Omega}

\def\pa{\partial}

\def\half{\frac{1}{2}}

\def\tr{{\rm tr}}

\def\Str{{\rm str}}
\def\and{{\rm and}}

\def\ie{{\it i.e.,} }

\def\IR{\mathbbm R}
\def\IZ{\mathbbm Z}

\def\A5{{AdS_5 \times S^5}}

\newcommand{\be}{\begin{equation}}
\newcommand{\bea}{\begin{eqnarray}}

\newcommand{\ee}{\end{equation}}
\newcommand{\eea}{\end{eqnarray}}

\begin{document}
\vspace*{-1.0in}
\thispagestyle{empty}
\begin{flushright}
CALT-TH-2020-004
\end{flushright}

%\large
\normalsize
\baselineskip = 18pt
\parskip = 6pt

\vspace{1.0in}

{\Large \begin{center}
{\bf The $\mathbf{AdS_5 \times S^5}$ Superstring}
\end{center}}

\vspace{.25in}

\begin{center}
John H. Schwarz\footnote{jhs@theory.caltech.edu}
\\
\emph{Walter Burke Institute for Theoretical Physics\\
California Institute of Technology 452-48\\ Pasadena, CA  91125, USA}
\end{center}
\vspace{.25in}

\begin{center}
\textbf{Abstract}
\end{center}
\begin{quotation}

The duality between the type IIB superstring theory in an $AdS_5 \times S^5$ background
with $N$ units of five-form flux
and ${\cal N} =4$ super Yang--Mills theory with a $U(N)$ gauge group has been studied
extensively. My version of the construction of the superstring world-sheet action is
reviewed here. This paper is dedicated to Michael Duff on the occasion of his 70th
birthday.

\end{quotation}

\newpage

\pagenumbering{arabic}

%\tableofcontents

%\newpage

\section{Introduction}

I am pleased to contribute to this volume honoring my good friend Michael Duff on the occasion
of his 70th birthday. Although we never collaborated, we have had lively scientific discussions, and
each of us has influenced the other's research. In the early 1980s, when Michael Green
and I were developing superstring theory, extra dimensions were not yet fashionable.
However, this began to change as a result of the work of Michael Duff and his collaborators
exploring compactifications of 11-dimensional supergravity \cite{Duff:1986hr}.
This work attracted a community of followers who acquired expertise that would prove useful in the
subsequent development of superstring theory and M theory.

Superstring theory was originally developed for a Minkowski spacetime geometry.
(For reviews see \cite{Green:1987sp}\cite{Polchinski:1998rq}\cite{Becker:2007zj}.)
This is the easiest case to handle mathematically, but
it is not the only possibility. In particular, there is great interest in Anti de Sitter (AdS)
geometries for studies of AdS/CFT duality.
This manuscript will review the specific case of the type IIB superstring in an $AdS_5 \times S^5$
background with $N$ units of five-form flux, which is dual to
${\cal N} =4$ super Yang--Mills theory with a $U(N)$
gauge group \cite{Maldacena:1997re}.\footnote{Since the $U(1)$ subgroup of $U(N)$ decouples, only
$SU(N)$ is relevant to our considerations. Nevertheless, $U(N)$, which
arises from $N$ coincident D3-branes, is the correct gauge group.}
This is an especially interesting example -- sometimes referred to as the ``hydrogen atom" of AdS/CFT.
It has a large group of symmetries. In particular, it is maximally supersymmetric (32 supercharges).
In addition, there are two tunable dimensionless parameters: the string coupling constant and the
ratio between the curvature radius and the string length scale. These are related by AdS/CFT duality
to the Yang--Mills coupling constant and the rank of the gauge group.

% okay to here.

In the Minkowski spacetime setting,
the free superstring spectrum was identified and shown not to contain ghosts or tachyons
in the critical spacetime dimension, which is 10 (nine space and one time).
Scattering amplitudes for $n$ massless external
on-shell particles can be constructed perturbatively in the string coupling constant $g_s$.
The most fundamental examples are the type IIA and type IIB superstring theories, which
only involve closed oriented strings and have maximal supersymmetry. For these theories
the massless states comprise a supergravity multiplet, and the conserved supercharges
consist of two Majorana--Weyl spinors, each of which has 16 real components. In the
type IIA case the two spinors have opposite chirality (or handedness), and the theory is parity
conserving. In the type IIB case they have the same chirality, and the theory
is parity violating. It is a nontrivial fact that the type IIB theory has no gravitational
anomalies \cite{AlvarezGaume:1983ig}.

String theories have a fundamental length scale, $l_s$, called the string scale. In units with
$\hbar = c=1$, one also defines the ``Regge-slope parameter" $\a' =l_s^2$ and the
fundamental string tension $T = (2\pi \a')^{-1}$. The string coupling constant is determined by the vacuum value of a massless scalar field $\phi$ , called the dilaton, $g_s = \langle e^\phi \rangle$.
$n$-particle on-shell scattering amplitudes for
both type II theories have a single Feynman diagram at each order of the perturbation expansion.
%\footnote{This remarkable fact has motivated the search for ways to simplify the calculation of
%field theory scattering amplitudes.}
At $g$ loops the unique string theory Feynman diagram
is a genus $g$ Riemann surface with $n$ punctures associated to the
external particles. This two-dimensional manifold is a Euclideanized
string world sheet. The $g$-loop amplitude is then given by an integral over the $3g+n-3$
complex-dimensional moduli space of such punctured Riemann surfaces.
These amplitudes are free of UV divergences.

There are two basic approaches to incorporating the fermionic
degrees of freedom.\footnote{There are also other approaches,
such as Berkovits' ``pure spinor" formalism.}
The first one, called the RNS formalism, involves
fermionic (\ie Grassmann valued) world-sheet fields that transform as
world-sheet spinors and spacetime vectors.
%In this approach it is necessary to include sums over various choices of ``spin structures." This
%concerns the choice of periodicity or antiperiodicity around various noncontractible cycles.
%Correct choices are required to implement spacetime supersymmetry.
%Extension of the Riemann surfaces to super Riemann surfaces (with Grassmann-odd dimensions
%and moduli) is natural in this approach.
One of the shortcomings of the RNS formalism is that nonzero backgrounds for fields
belonging to the RR sector are difficult to incorporate. The second basic approach,
called the GS formalism, utilizes fermionic world-sheet fields that transform as world-sheet
scalars and spacetime spinors.
%Since there are no world-sheet spinors, this approach does not utilize sums over
%spin structures or the introduction of super Riemann surfaces.
It can handle background RR fields and it makes spacetime supersymmetry manifest. However,
the rules for constructing multiloop amplitudes have not been worked out in the GS formalism.
This paper will describe type IIB superstring theory in an $AdS_5 \times S^5$
background. Since this background includes a nonzero RR field, a five-form field strength,
the GS formalism is best suited to this problem.

%\section{AdS/CFT duality}

Maldacena's original paper proposing AdS/CFT duality drew attention to three (previously known)
maximally supersymmetric geometries containing $AdS$ factors \cite{Maldacena:1997re}.
${AdS_4 \times S^7}$ and ${AdS_7 \times S^4}$ are M-theory backgrounds, whereas
${AdS_5 \times S^5}$ is a type IIB superstring theory background. The latter case can be
studied in greatest detail, because superstring theory is better understood than M theory,
which does not have a dimensionless coupling constant. The dual conformal field theory (CFT)
in this case is four-dimensional ${\cal N} =4$ super Yang--Mills theory
with a $U(N)$ gauge group. The integer parameter $N$ corresponds to the amount of self-dual five-form
flux, the nonzero RR field, in the 10d geometry.

The goal of this paper is to describe the world-sheet action of a type IIB superstring
in the ${AdS_5 \times S^5}$ background geometry, with the appropriate self-dual five-form flux.
The isometry group of type IIB superstring theory in this background
is given by the supergroup $PSU(2,2|4)$. This problem was originally studied in
the GS formalism \cite{Metsaev:1998it}\cite{Kallosh:1998nx}\cite{Kallosh:1998ji}\cite{Roiban:2000yy}
using a formulation based on the quotient space
$ PSU(2,2|4)/SO(4,1) \times SO(5)$. This paper will summarize closely related much later work
using a slightly different approach \cite{Schwarz:2015lla}. One advantage of this approach
is that it is based on world-sheet
supermatrices whose full dependence on the ten bosonic coordinates and the
32 Grassmann coordinates is completely explicit. I am optimistic that there are additional
advantages, but that has not yet been demonstrated.
We will also review the proof that this action describes an integrable
theory \cite{Bena:2003wd}\cite{Arutyunov:2004yx}\cite{Alday:2005gi}\cite{Beisert:2010jr}.
On the other hand, this short review shall omit a couple of important technical issues.
One is the demonstration that the theory has a $\IZ_4$ symmetry. In other approaches it
is utilized to prove integrability. We shall establish integrability without invoking that symmetry.
Another omitted technicality is the proof of kappa symmetry.

\subsection{The bosonic truncation}

The unit-radius sphere $S^5$ can be described as a submanifold of $\IR^6$
\be
 \hat z \cdot \hat z = (z^1)^2 + (z^2)^2 + \ldots + (z^6)^2 = 1.
\ee
Similarly, the unit-radius anti de Sitter space $AdS_5$ can be described
by\footnote{Strictly speaking, this describes the Poincar\'e patch
of $AdS_5$.}
\be
 \hat y \cdot \hat y = -(y^0)^2 + (y^1)^2 + \ldots +(y^4)^2 - (y^5)^2 = -1,
\ee
which is a submanifold of $\IR^{4,2}$.
These formulas make the symmetries $SO(6)$ and $SO(4,2)$, respectively, manifest.
When we add fermions these groups will be replaced by their covering groups, which are
$SU(4)$ and $SU(2,2)$.
In this notation the ${AdS_5 \times S^5}$ metric of radius $R$ takes the form
\be
 ds^2 = R^2(d\hat y \cdot d \hat y + d\hat z \cdot d \hat z).
\ee

The induced world-volume metric of a probe $p$-brane with local coordinates
$\s^\a$, $\a=0,1,\ldots,p$ is
\be
 G_{\a\b} (\s) = \pa_\a \hat z(\s) \cdot \pa_\b \hat z(\s)
 + \pa_\a \hat y(\s) \cdot \pa_\b \hat y(\s) .
\ee
As usual, the vector functions $\hat z(\s)$ and $\hat y (\s)$ describe the
spacetime embedding of the brane. In this work we are concerned with the superstring
for which $p =1$.
The bosonic part of the radius $R$ superstring action can be written in the general
coordinate invariant form
\be
 S = - \frac{R^2}{4 \pi \a'} \int d^2 \s \sqrt{-h} h^{\a\b} G_{\a\b},
\ee
where $h_{\a \b}(\s)$ is an auxiliary world-sheet metric.
This metric is related to the induced metric
$G_{\a\b}(\s)$, up to a Weyl rescaling,
by the $h$ equation of motion. The AdS/CFT dictionary implies that
\be
R^2 = \a'\sqrt{\l} \quad {\rm and} \quad g_{YM}^2 = 4\pi g_s,
\ee
where $\l = g_{YM}^2 N$ is the 't Hooft parameter of the dual CFT, which
is ${\cal N} =4$ super Yang--Mills theory with gauge group $U(N)$.
%Also, $\a'$ is the Regge-slope parameter,
%which (as in flat space) is related to the fundamental string tension $T$ by $2\pi \a' T = 1$.

\section{Supermatrices and supergeometry}

%\subsection{Supermatrices}

In order to add fermionic degrees of freedom, it is convenient to introduce Grassmann
coordinates. Towards this end, let us first discuss supermatrices,
which we write in the form\footnote{Other authors use slightly different, but equivalent,
conventions.}
\be
S
= \left( \begin{array}{cc}
a &  \z b  \\
\z c & d \\
\end{array} \right), \quad  \z =  e^{-i\pi/4}
\ee
$a$ and $d$ are $4 \times 4$ matrices  of Grassmann-even numbers  referring to
$SU(4)$ and $SU(2,2)$. On the other hand, $b$ and $c$
are $4 \times 4$ matrices of Grassmann-odd numbers that transform as bifundamentals,
$(\mathbf{4, \bar 4})$ and $(\mathbf{\bar 4, 4})$, of the two groups.

The ``superadjoint'' is defined by
\be
S^{\dagger}
= \left( \begin{array}{cc}
a^{\dagger} &  -\z c^{\dagger}  \\
-\z b^{\dagger} & d^{\dagger} \\
\end{array} \right) .
\ee
Using the rule $(cb)^{\dagger} = - c^{\dagger} b^{\dagger}$,
this definition ensures that $(S_1 S_2)^{\dagger} = S_2^{\dagger} S_1^{\dagger}$
and $(S^{\dagger})^{\dagger}=S$, as one can easily verify. By definition,
a unitary supermatrix satisfies $S S^{\dagger} = I$, where $I$ is the unit matrix,
and an antihermitian supermatrix satisfies $S + S^{\dagger} = 0$.\footnote{Additional
minus signs, needed to take account of the indefinite signature
of $SU(2,2)$ are suppressed in this discussion.} In this way one
defines the super Lie group $SU(2,2|4)$ and the super Lie algebra $\mathfrak{su}(2,2|4)$.
The ``supertrace'' is defined (as usual) by
\be
\Str S = \tr\, a - \tr \, d .
\ee
The main virtue of this definition is that
\be
 \Str (S_1 S_2) = \Str (S_2 S_1).
\ee
Note also that $\Str \, I = 0$, since the $a$ and $d$ blocks are both $4 \times 4$.
%The {\em supertranspose} is defined by
%\begin{equation} \label{supertranspose}
%M^T
%= \left( \begin{array}{cc}
%a^T &  - i\z c^T  \\
%-i\z b^T & d^T \\
%\end{array} \right) .
%\end{equation}
%This definition ensures that $ (M_1 M_2)^T = M_2^T M_1^T$. However, it has the somewhat
%surprising property
%\begin{equation}
%(M^T)^T =  \left( \begin{array}{cc}
%a &  -\z b  \\
%-\z c & d \\
%$\end{array} \right) ,
%$\end{equation}
%which makes the supertranspose a $\IZ_4$ operation.
The $\mathfrak{psu}(2,2|4)$ algebra, which is the one that is required,
does not have a supermatrix realization. Rather, it is described by
$\mathfrak{su}(2,2|4)$ supermatrices modded out by the equivalence
relation
\be
S \sim S + \l I.
\ee

%\subsection{Nonlinear realization of the superalgebra}

In addition to the bosonic $y$ and $z$ coordinates, described in the introduction,
we will utilize $\th$ coordinates, which are 16 complex Grassmann numbers
that transform under $SU(4)\times SU(2,2)$ as ${\bf (4, \bar4)}$.
It is natural to describe them by $4 \times 4$ matrices, rather
than by $32$-component spinors as is traditionally done for the flat-space theory.
This has the advantage that no Fierz transformations are required
in the analysis. Infinitesimal symmetry transformations are given by the rule
\be
 \d\th = \o\th - \th \tilde\o +\e + i \th\ed \th ,
\ee
where $\o$ belongs to the $\mathfrak{su}(2,2)$, $\tilde\o$ belongs
to $\mathfrak{su}(4)$, and $\e$ is a bifundamental matrix of Grassmann numbers.
It is straightforward to verify that iteration of this formula
closes on the $\mathfrak{psu}(2,2|4)$ algebra.
It is reminiscent of Goldstino transformations in theories with
spontaneously broken supersymmetry.

Supermatrices $\G(\th) \in SU(2,2|4)$ can be written in the form
\be
 \G (\th) = \left( \begin{array}{cc}
I &  \z \th  \\
\z \td & I \\
\end{array} \right)  \left( \begin{array}{cc}
f^{-1} &  0  \\
0 & {\tilde f}^{-1} \\
\end{array} \right)
\ee
by choosing $f$ and $\tilde f$ such that $\G \G^{\dagger} = I$.
This is achieved for
\be
f = \sqrt{I+u} = I + \half u + \ldots
\ee
\be
\tilde f = \sqrt{I + \tilde u} = I + \half \tilde u +\ldots,
\ee
where
\be
 u = i \th \td \quad {\rm and} \quad \tilde u = i \td\th .
 \ee
It then follows that
\be
\d_{\e}\G  = \G \left( \begin{array}{cc}
M(\e) &  0  \\
0 & \tilde M(\e) \\
\end{array} \right)  +   \left( \begin{array}{cc}
0 &  \z \e  \\
\z \ed & 0 \\
\end{array} \right)\G,
\ee
where
\be
 M(\e) = -(\d_{\e} f - i f \e \td) f^{-1},
\ee
\be
 \tilde M(\e) = -(\d_{\e} \tilde f - i \tilde f \ed \th) \tilde f^{-1}.
 \ee
%The natural interpretation is that $\G(\th)$ describes the coset space
%\be
%PSU(2,2|4)/ SU(4) \times SU(2,2).
%\ee
Now consider the superconnection
\be \label{Adef}
 {\cal A} = \G^{-1} d\G
= \left( \begin{array}{cc}
K &  \z \Psi  \\
\z \Psi^{\dagger} & \tilde K \\
\end{array} \right).
\ee
This one-form supermatrix, constructed entirely out of $\th$,
is super-antihermitian and flat ($d{\cal A} + {\cal A}\wedge {\cal A} =0$). This formula
defines the even one-forms $K$ and $\tilde K$, and the
odd one-forms $\Psi$ and $\Psi^{\dagger}$.
Under a supersymmetry transformation
\be
 \d_{\e}{\cal A} = - d\left( \begin{array}{cc}
M &  0  \\
0 & \tilde M \\
\end{array} \right) -  \left[ {\cal A} , \left( \begin{array}{cc}
M &  0  \\
0 & \tilde M \\
\end{array} \right) \right],
\ee
where $M$ and $\tilde M$ are the matrix functions defined above.

%okay to here

Let us now recast the bosonic coordinates $\hat y$ and $\hat z$,
defined in the introduction, in matrix form.
$S^5$ can be represented by the antisymmetric $SU(4)$ matrix:
\be
Z = \left( \begin{array}{cccc}
0 & u & v & w  \\
-u & 0 & -\bar w & \bar v\\
-v & \bar w & 0 & -\bar u\\
- w & - \bar v & \bar u & 0 \\
\end{array} \right)  = \S_a z^a,
\ee
where $u = z^1 +i z^2$, $v = z^3 + i z^4$, and $w = z^5 + iz^6$.
Requiring $|u|^2 + |v|^2 + |w|^2 =1$, it then follows that
\be
 Z = - Z^T, \quad Z Z^{\dagger} = I, \quad \det Z =1 .
\ee
The main purpose in displaying all the elements of the matrix $Z$
is to establish the existence of a matrix with these properties.
The matrix $Z$ defines a codimension 10 map of $S^5$ into $SU(4)$.
There is a very similar construction for $Y: AdS_5 \to SU(2,2)$.

The supersymmetry transformations of the bosonic coordinates are
\be
 \d_\e Z = M Z + Z M^T \quad {\rm and} \quad
\d_\e Y = \tilde M Y + Y {\tilde M}^T.
\ee
It then follows that the antihermitian connections
\be
 {\O}  = Z d \Zd -  K - Z { K}^T \Zd,
\ee
\be
{\tilde\O}  = Y d \Yd - \tilde K - Y {\tilde K}^T \Yd
\ee
transform nicely under supersymmetry transformations:
\be
\d_{\e} \O = [M, \O]  \quad {\rm and} \quad
\d_{\e} \tilde\O = [\tilde M, \tilde\O].
\ee
The $PSU(2,2|4)$ invariant metric with the
correct bosonic truncation is
\be
G_{\a\b} = -\frac{1}{4} \left(\tr(\O_{\a} \O_{\b}) - \tr(\tilde\O_{\a}\tilde\O_{\b}) \right).
\ee

The next step is to split the Grassmann-odd matrix $\Psi$
in the superconnection ${\cal A}$, transforming as
$({\bf 4, \bar 4})$ under $SU(4) \times SU(2,2)$, into two pieces that
correspond to Majorana--Weyl (MW) spinors
while respecting the group theory. To do this, we define an involution
\be
\Psi \to \Psi' = Z \Psi^{\star} \Yd.
\ee
$\Psi'$ also transforms as $({\bf 4, \bar 4})$. Then we can write
\be
\Psi = \Psi_1 + i \Psi_2  \quad {\rm and} \quad \Psi' = \Psi_1 -i \Psi_2,
\ee
where $\Psi_1$ and $\Psi_2$ are ``MW matrices'' for which $\Psi_I' = \Psi_I,$
$I=1,2$.

Let us now define three antihermitian supermatrix one-forms out of quantities
introduced above:
\be
A_1
= \left( \begin{array}{cc}
\O & 0  \\
0 & \tilde\O \\
\end{array} \right) ,
\quad
A_2
= \left( \begin{array}{cc}
0 &  \z \Psi  \\
\z \Psi^{\dagger} & 0 \\
\end{array} \right) ,
\quad
A_3
= \left( \begin{array}{cc}
0 &  \z \Psi'  \\
\z \Psi^{\prime\dagger} & 0 \\
\end{array} \right).
\ee
In all three cases, infinitesimal supersymmetry transformations take the form
\be \d_{\e}A_i =
 \left[\left( \begin{array}{cc}
M &  0  \\
0 & \tilde M \\
\end{array} \right), A_i \right] \quad i = 1,2,3.
\ee
Next, we define
\be
 J_i = \G^{-1} A_i \G \quad i=1,2,3,
\ee
to obtain supermatrices that transform under $\mathfrak{psu}(2,2|4)$ in the natural way, namely
\be
 \d_{\L} J_i = [ \L , J_i ],
\ee
where the infinitesimal parameters are incorporated in the supermatrix
\be
\L = \left( \begin{array}{cc}
\o &  \z \e  \\
\z \e^{\dagger} & \tilde\o \\
\end{array} \right) .
\ee
We will formulate the superstring action and its equations of motion
in terms of the three one-forms $J_i$.
Using the explicit formulas given above, one obtains the
Maurer--Cartan equations\footnote{The relation to the notation of \cite{Bena:2003wd}
is $J_1 = 2p$, $J_2 = q$, $J_3 = q'$.}
\be
dJ_1 = -J_1 \wedge J_1 + J_2 \wedge J_2 +J_3 \wedge J_3 -J_1 \wedge J_2 -J_2 \wedge J_1,
\ee
\be
dJ_2 = - 2 J_2 \wedge J_2,
\ee
\be
dJ_3 = -(J_1 + J_2)\wedge J_3
-J_3 \wedge (J_1 + J_2).
\ee

%okay to here.

\section{The superstring world-sheet theory}

The superstring action contains two pieces: the metric term discussed previously and
a Wess--Zumino (WZ) term.
The WZ term for the fundamental string can be expressed in terms of a $\mathfrak{psu}(2,2|4)$
invariant two-form. Candidate invariant two-forms are
\be
\Str(J_i \wedge J_j)= - \Str(J_j \wedge J_i).
\ee
However, the only one of these that is nonzero is $\Str( J_2 \wedge J_3)$.
Therefore the WZ term of the fundamental string is proportional to
$\int \Str( J_2 \wedge J_3)$.
%\footnote{This two-form also enters in the DBI term
%of the D3-brane action, where it is added to the Maxwell field strength $F=dA$.}
This term does not contribute to the bosonic truncation.

The supersymmetric extension of the induced
world-volume metric in the introduction can now be recast in the form
\be
G_{\a\b} =  -\frac{1}{4}\Str(J_{1\a} J_{1\b}).
\ee
The complete superstring world-sheet action in the $AdS_5 \times S^5$ background is then
\be
 S =  \frac{\sqrt{\l}}{16\pi}\int \Str(J_1 \wedge \star J_1)
- \frac{\sqrt{\l}}{8\pi} \int\Str(J_2 \wedge J_3).
\ee
This action is manifestly invariant under $\d_{\L} J_i = [ \L , J_i ]$, and thus it has $PSU(2,2|4)$ symmetry.
It allows one to deduce that the $PSU(2,2|4)$ Noether current is
\be
J_N = J_1 + \star J_3.
\ee
The conservation of this current encodes all of the superstring equations of motion,
which take the remarkably concise form
\be
 d \star J_N = d \star J_1 + d J_3 =0.
\ee
Even so, they are fairly complicated when written out in term of the matrices $Z$, $Y$, and $\th$.

This theory resembles a sigma model for a symmetric space. However, it has two features that
are specific to string theory. The first is the fact
that the Hodge duals ($\star J$) in the preceding paragraph are defined using the
auxiliary metric $h_{\a\b}$. As a consequence, the theory has reparametrization invariance and
Weyl symmetry, which are local symmetries. The second feature
is a local fermionic symmetry, called kappa symmetry. It determines the ratio
of the coefficients of the two terms in the action (up to a sign).
There are well-known ways of analyzing the consequences of these features for the flat
space ($R \to \infty$) limit of this theory, which generalize to the present setting.
Reparametrization invariance of the world-sheet theory allows one to choose a gauge
in which $h_{\a\b}$ is conformally flat, \ie $h_{\a\b}(\s) = \L (\s) \eta_{\a\b}$,
where $\eta$ is the 2d Minkowski metric. In critical string theories the decoupling of the
conformal factor $\L$, \ie the Weyl symmetry,
remains valid at the quantum level. This implies that the trace of the
stress tensor $T_{\a\b}$ vanishes. In world-sheet light-cone coordinates ($\s^{\pm} = \s^0 \pm \s^1$),
the remaining $h_{\a\b}$ equations of motion are the stress tensor constraints
$T_{++} = T_{--} =0$, which
amount to $\Str (J_{1+} J_{1+}) = \Str (J_{1-} J_{1-}) =0$. In the quantum treatment, these operators
are required to have vanishing matrix elements between physical states. Also, local kappa symmetry
allows one to choose a gauge in which half of the $\th$ coordinates are set
to zero. The flat-space limit should give the type IIB
superstring world-sheet theory in 10d Minkowski spacetime, which is a free theory in an appropriate
gauge \cite{Green:1983wt}.

The $AdS_5 \times S^5$ superstring theory is not a free theory, but it is integrable.
The key to the proof of this fact is to establish that there is
a one-parameter family of flat connections $J(t)$,
\be
dJ(t) + J(t) \wedge J(t) =0.
\ee
Given a flat connection $J(t)$,
there is a standard procedure for constructing an infinite family of conserved charges and
arguing that this implies integrability \cite{Luscher:1977rq}\cite{Brezin:1979am}.

Before constructing $J(t)$ for the superstring action, let us
first consider its bosonic truncation. Classically, integrability of the supersymmetric
theory requires integrability of the bosonic truncation. At the quantum level, the supersymmetric theory
is expected to be better behaved than its bosonic truncation, because it is in its critical
dimension.\footnote{The quantum theory of this 2d string action
describes the classical behavior of the spacetime string in 10d. In other words, it gives the full
$ \sqrt{\l} = R^2/\a' $ dependence at leading order in $ g_s $.}
In the bosonic truncation, we have $J_2 = J_3 =0$ and $J_1$ is flat: $dJ_1 = - J_1 \wedge J_1$.
Moreover, the Noether equations (and the equations of motion) are simply $d \star J_1 =0$.
%in fact this is the structure of
%a symmetric-space theory, which in this case is based on the coset
%\be
%\frac{SO(6)}{SO(5)} \times \frac{SO(4,2)}{SO(4,1)}.
%\ee
Let us now consider linear combinations of the form
\be
J(t) = c_1(t) J_1 + d_1(t) \star J_1.
\ee
This is a flat connection provided that $c_1 = (c_1)^2 - (d_1)^2$,
which is solved for the one-parameter family of choices
\be
c_1(t) = - \sinh^2 t, \quad d_1(t) =  \sinh t \cosh t.
\ee

Let us now generalize the preceding to include the fermionic degrees of freedom.
The Maurer--Cartan equations, together with $d\star J_1 + dJ_3 =0$,
imply that
\be
 J(t) = c_1(t) J_1 + d_1(t)\star J_1 + c_2(t) J_2 + c_3(t) J_3
\ee
is flat for a one-parameter family given by\footnote{The alternative choices $c_1'=c_1$,
$d_1' = -d_1$, $c_2' = 2 - c_2$, $c_3' =c_3$ in \cite{Schwarz:2015lla} and \cite{Bena:2003wd}
are obtained from these by $t \to i\pi -t$.}
\be
c_2(t) = 1 - \cosh t, \quad c_3(t) = \sinh t
\ee
together with the previous choices of $c_1(t)$ and $d_1(t)$ \cite{Bena:2003wd}.
To verify this result it is useful to know that the equations of motion imply that \cite{Schwarz:2015lla}
\be
\star J_1 \wedge J_2 + J_2 \wedge \star J_1 = - (J_1 \wedge J_3 + J_3 \wedge J_1),
\ee
\be
\star J_1 \wedge J_3 + J_3 \wedge \star J_1 = - (J_1 \wedge J_2 + J_2 \wedge J_1).
\ee

Each of the two terms in the superstring action has manifest $PSU(2,2|4)$
global symmetry and reparametrization invariance. As mentioned earlier,
the requirement of kappa symmetry determines the ratio
of the coefficients of the two terms up to a sign. Fortunately, this ratio is the same as the one
required for integrability.

%okay to here.

\section{Conclusion}

The $AdS_5 \times S^5$ superstring action
described here exhibits the complete $\th$ dependence of all quantities,
and it has manifest $PSU(2,2|4)$ symmetry. In contrast to its flat-space
limit, this is an interacting world-sheet theory, so it is much more challenging
to give a complete quantum description of its spectrum and other properties.
Nevertheless, it has been studied in great detail (using earlier formulations)
and compared, with remarkable success, to the dual CFT by taking advantage of
the fact that it is integrable \cite{Beisert:2010jr}.

\section{Funding statement}

This research has been supported in part by the Walter Burke Institute
for Theoretical Physics at Caltech and by U.S. DOE Grant DE-SC0011632,
and some of this work was performed at the Aspen Center for Physics,
which is supported by National Science Foundation grant PHY-1607611.

\bigskip

%okay to here

\end{document}